\documentclass[conference]{IEEEtran}
\IEEEoverridecommandlockouts
\usepackage{cite}
\usepackage{amsmath,amssymb,amsfonts}
\usepackage{algorithmic}
\usepackage{graphicx}
\usepackage{textcomp}
\usepackage{xcolor}
\def\BibTeX{{\rm B\kern-.05em{\sc i\kern-.025em b}\kern-.08em
    T\kern-.1667em\lower.7ex\hbox{E}\kern-.125emX}}

\usepackage{color}

\usepackage{mathtools}
\usepackage{booktabs}
\usepackage{verbatim}

\usepackage{varwidth}
\usepackage{multirow}
\usepackage{rotating, xcolor}
\usepackage{colortbl}
\usepackage{caption}
\usepackage[subrefformat=parens,labelformat=parens]{subcaption}
\usepackage{enumitem}
\usepackage[hyphens]{url}
\usepackage{float}
\floatstyle{ruled}
\usepackage{tcolorbox}
\definecolor{mycolor}{rgb}{0.122, 0.435, 0.698}


\usepackage{mathtools}

\title{Introducing the Quantum Research Kernels: Lessons from Classical Parallel Computing 
}

\author{\IEEEauthorblockN{A. Y. Matsuura}
\IEEEauthorblockA{\textit{Applications and Quantum Architecture Lab} \\
\textit{Intel Corp.}\\
Hillsboro, OR, USA \\
anne.y.matsuura@intel.com}
\and
\IEEEauthorblockN{Timothy G.  Mattson}
\IEEEauthorblockA{\textit{Parallel Computing Lab} \\
\textit{Intel Corp.}\\
Ilwaco, WA, USA \\
timothy.g.mattson@intel.com}
}

\begin{document}
\maketitle

\begin{abstract}

Quantum computing represents a paradigm shift for computation requiring an entirely new computer architecture. However, there is much that can be learned from
traditional classical computer engineering. In this paper, we describe the Parallel Research Kernels (PRK), a tool that was very useful for designing classical
parallel computing systems. The PRK are simple kernels written to expose bottlenecks that limit classical parallel computing performance. We hypothesize that an
analogous tool for quantum computing, Quantum Research Kernels (QRK), may similarly aid the codesign of software and hardware for quantum computing systems, and 
we give a few examples of representative QRKs.

\end{abstract}


\section{Introduction}
\label{sec:intro}

Quantum computers have the potential to transform computing. Their advent promises to open up new classes of applications and enable the solution of problems that are intractable for
classical computers today.  To reach that potential, however, requires great technological innovations, including research advances in fundamental physics, the
development of new quantum programming models that ``normal'' humans can use, and solutions to a host of issues in both software and hardware concerning building quantum
systems.

Quantum computing has a long way to go, but we believe we can leverage best practices from classical digital computing to advance through the stages of development
at a faster rate.   In this paper, we focus on one lesson learned from the era of digital computing: that the best systems emerge from a hardware/software co-design
process.   All too often hardware is designed and then ``thrown over the fence'' to software developers to figure out how to 
make the hardware useful.   It is more effective to have the software and hardware teams working together, so
the programming systems are ready as soon as the hardware is available and the hardware includes features
specifically needed to make the software both more efficient to run and easier to write.

However, it is challenging to do hardware/software codesign when you don't know what future applications will look like.    Hypothesis: Even though we 
may not know what future applications will look like, we do know the features of a system that will limit these applications.
If we can define these ``bottlenecks'' and build a system that collectively minimizes their impact, we can be confident that the 
system will be effective for these future applications.   

This was the basic principle behind Parallel Research Kernels (PRK) ~\cite{PRK2014} in the parallel computing field.     The PRK are a collection of simple kernels
that expose the features of a system that limits parallel performance.  They are small, generate their own data, do a computation (so
systems can't ``cheat''), and test their results.  In essence, they are a way for application programmers to precisely define what they need
from a system; to guide hardware developers to build systems that will work well for applications.
They have proven useful for designing systems~\cite{SCC10} and to explore the suitability of extreme scale programming models~\cite{prkexa:16}.

The goal of this paper is to launch a conversation about using an approach similar to the PRK for quantum computing.  We call these the
Quantum Research Kernels (QRK).   Can we define features of quantum computing systems that will limit applications for these systems?
Can we produce well defined kernels to expose these features?   Can we anticipate the breadth of application design patterns so we can
be confident that the set of QRK are complete?    These are challenging questions.  In this paper, we define the problem 
and propose a process to answer these questions.

\section{Parallel Research Kernels}
\label{sec:prk}

The Parallel Research Kernels stress a system in ways parallel applications in high performance computing would.
They are listed in Table~\ref{table:PRK} where we provide the name of each kernel, a brief definition, and 
the features of a parallel system stressed by the kernel.  The kernels are defined mathematically and with a 
reference implementation using C, OpenMP, and MPI.  They are available in a github repository~\cite{prkRepo}.

\begin{table}[!htbp]
\centering
\caption{\textbf{Parallel Research Kernels:} 
-- \small
A set of basic kernels designed to stress features of a system that limit the performance of HPC applications.}
\label{table:PRK}
\begin{tabular}{|l|l|l|}
\hline
\emph{Name}   &   \emph{Definition}                  & \emph{Exposed system feature}                     \\                    
\hline

Transpose        &  Transpose a dense matrix     &   Bisection Bandwidth                      \\
\hline

Reduce            &  Elementwise sum of               &   Message passing, local           \\
                        &  multiple private vectors           &   memory bandwidth                  \\
\hline

Sparse             &  Sparse-martix vector product &  scatter/gather operations      \\
\hline

Random             & Random update to a table    &  Bandwidth to memory  \\
                            &                                             &   with random updates   \\
\hline

Synch\_global  &  Global synchronization           &    Collective synch.       \\
\hline

Synch\_p2p      &  point-to-point                         &  message passing latency,   \\
                         & synchronization                       &  remote atomics  \\
\hline

Stencil              & Stencil method                         &  nearest neighbor and   \\
                         &                                                  & asynchronous comm.       \\
\hline

Refcount          &  Update shared or private         & Mutual exclusion locks                        \\
                        &  counters                                   &                                                            \\
\hline

Nstream          &  daxpy over large vectors          & Peak memory bandwidth                  \\
\hline

DGEMM          & Dense Matrix Product               &  Peak floating point perf.      \\
\hline

Branch           & Inner loop with branches             &  Misses to the instruction     \\
                       &                                                     & cache, branch prediction.    \\
\hline

PIC                  & Particle in cell                           &   unstructured asynch. \\
                         &                                                  &multitasking \\
\hline

AMR                & Adaptive mesh refinement &   hierarchical asynch.    \\
                         &                                                  &multitasking \\
\hline
\end{tabular}
\end{table}

We submit that the Parallel Research Kernels are complete.  If a system is built that does well with all of them, then
it is  likely that system will be effective for running parallel HPC applications.  The PRK were selected by an ad hoc committee
of parallel application programmers.  When we started the project (in 2005) parallel computing in various forms had been around 
over 25 years.  We were able to convene a committee of experts with decades of experience.  Over the course of several meetings
and long email chains, the committee came up with the list of kernels.

\section{Quantum Research Kernels}
\label{sec:qrk}

The quantum research kernels (QRK), inspired by the PRK, will define a set of kernels designed to 
expose features of a quantum computing system that will constrain the success of applications
written for quantum computers.  The QRK are intended to be a sufficiently complete set such that
if a system were constructed that did well for each of the QRK, that system would most likely be
a successful system for supporting key applications.

As with the PRK, the QRK will be produced through a community-driven process by programmers interested in
writing applications for quantum computers.   To help nurture this conversation, we provide an example
of a potential QRK. One of the bottlenecks for scalable quantum computing, is the 
difficulty of loading the data (particularly classical data) into the quantum
machine. State preparation can be an exponentially hard problem itself, leading
to the conundrum that loading of the data into the quantum machine can completely
negate the speed-up gained by doing the problem on a quantum computer \cite{aronson}.

Hence, the first QRK follows.
\begin{itemize}
\item {\bf Name}: Encode
\item{\bf Definition}: Create a sequence of classical values from $i=0$ to $i=N$ equal to $4i{\pi}/N$
\item{\bf Action}: Encode qubits with the values from the previous step.   Rotate each qubit by ${\pi}/6$.
\item{\bf Test}: Read qubits and confirm that they have the correct rotated value
\end{itemize}

Note this has all the features we would expect in a QRK.  It defines problem input that is generated so the QRK can 
scale to any number of qubits.  A specific operation is defined so at the end of the QRK, the result can be validated.  
Finally, it exposes a specific feature of a quantum computer that will limit the ability of applications to run on the system
expressed in terms that architects can understand and use to guide the design of a quantum computer.

We have the beginnings of two additional QRK.  The first is one we call the \emph{Computational Area}.  
This is the product of a number of qubits that can be entangled times the number of operations that can be
carried out before the entangled state can no longer be maintained.  A high Computational Area can be produced by a small number of 
qubits that remain entangled over a large number of operations or by having a large number of qubits that remain entangled
for a small number of operations.    There are advantages to both cases so this measure supports both.

The second additional QRK is called \emph{Parallel Streams}.  This measures the ability of a quantum computer
to execute multiple independent streams of operations at the same time and in parallel. There are a number of options on how
to define the work that must be carried out in parallel.  Initially, we would run the Computational Area QRK in each stream, though
for a wider range of applications for quantum computers, we may find a better case for each stream.

These three QRK are just a start.  We need a larger set that covers the full range of features needed from a successful 
quantum computer, hopefully, on the order of 10.  The system features each QRK stresses overlap, but taken together
they need to cover the full range of features need by application programmers from a quantum computer so those designing
parallel systems can be confident a system that does well on the full set of QRK will meet the needs of applications programmers. 
This is the primary goal of the QRK.  However, 
once established, an application designer can model the needs of an application in terms of a linear combination of QRK
thereby using them to help select the quantum computer best suited to a particular application.

\section{Conclusion}
\label{sec:conclude}

We believe that the quantum computing field can learn from classical computing design techniques. The PRK are a powerful design tool, 
and we posit that there is a need for a similar approach for hardware/software co-design for quantum computing. We have introduced the concept
of QRK and have provided a few examples of QRK. This paper is a call to the quantum computing user community to work 
together to develop a complete set of QRK that can guide the design and development of quantum computing.

\bibliographystyle{IEEEtran}
\bibliography{QRK}

\end{document}